\documentclass[final,authoryear,5p]{elsarticle}

\usepackage{epsfig}

\usepackage{pdflscape} 

\usepackage{amssymb}
\usepackage{amsmath}

\usepackage{lineno}  

\usepackage[usenames]{color}

\usepackage[ps2pdf,%
a4paper=true,%
breaklinks=true,%
colorlinks=true,%
pdfauthor={First Author et al.},%
pdftitle={Template for manuscripts in Advances in Space Research}%
]{hyperref}

\bibpunct{(}{)}{;}{a}{}{,}  

\journal{Journal of High Energy Astrophysics}

\newcommand{\sw}{{\it Swift}}
\newcommand{\chandra}{{\it Chandra}}
\newcommand{\inte}{INTEGRAL}
\newcommand{\nustar}{{\it NuSTAR}}
\newcommand{\rxte}{{\it RXTE}}
\newcommand{\xmm}{{\it XMM--Newton}}

\def \aip {AIP Conf. Ser.}
\def \ATel {ATel} 
\def \apj {ApJ}
\def \apjl {ApJL}
\def \apjs {ApJS}
\def \aap {A\&A}

\def \asr {AdSpR}
\def \baas {BAAS}

\def \mnras {MNRAS}

\def \ssr {SSRv}

\begin{document}

\begin{frontmatter}

\title{Seven Years with the {\it Swift} Supergiant Fast  X-ray Transients Project\tnoteref{footnote1}}
\tnotetext[footnote1]{  \href{http://www.ifc.inaf.it/sfxt/}{Project web page: http://www.ifc.inaf.it/sfxt/ }}

\author[a]{P.\ Romano\corref{cor}} 
\ead{romano@ifc.inaf.it}

\address[a]{INAF-IASF Palermo,  Via U.\ La Malfa 153, I-90146 Palermo, Italy}
\cortext[cor]{Corresponding author}

\begin{abstract}
Supergiant Fast X-ray Transients (SFXTs) are HMXBs with OB supergiant companions.  
I review the results of the  {\it Swift} SFXT Project, which since 2007 has been exploiting {\it Swift}'s 
capabilities in a systematic study of SFXTs and supergiant X-ray binaries (SGXBs) 
by combining follow-ups of outbursts, when detailed broad-band spectroscopy is possible, 
with long-term monitoring campaigns, when the out-of-outburst fainter states can be observed. 
This strategy has led us to measure their duty cycles as a function of luminosity, 
to extract their differential luminosity distributions in the soft X-ray domain, and to 
compare, with unprecedented detail, the X-ray variability in these different classes of sources. 
I also discuss the ``seventh year crisis'', the challenges that the recent {\it Swift} observations  
are making to the prevailing models attempting to explain the SFXT behaviour. 
\end{abstract}

\end{frontmatter}

\parindent=0.5 cm


            \section{Introduction \label{swift10:intro}}

Supergiant X-ray binaries (SGXBs) hosting an accreting neutron star and an OB 
supergiant companion, are divided into 
{\it classical systems}, showing a strong X-ray variability with an X-ray luminosity dynamic range of 10--50, 
and {\it supergiant fast X-ray transients} \citep[SFXTs,][]{Smith2004:fast_transients,Sguera2005,Negueruela2006:ESASP604}.
The dozen or so members of the latter class \citep[see][for a recent review]{Romano2014:sfxts_catI}, 
have a quiescent luminosity of $\sim 10^{32}$~erg~s$^{-1}$ \citep[][]{zand2005,Bozzo2010:quiesc1739n08408}
and display X-ray flares reaching 10$^{36}$--10$^{37}$~erg~s$^{-1}$. 
They are therefore identified based on their characteristic high dynamic range in X-ray luminosity, 
which reaches up to $\sim$10$^3$--10$^5$ times the range observed in 
classical systems \citep[][]{Sguera2005,Romano2015:17544sb}, even though 
the supergiant stars in SFXTs and classical SGXBs share 
similar orbital periods and spectroscopic properties. 
The origin of this different behaviour is still a matter of debate 
\citep[see, e.g.][]{Bozzo2013:COSPAR_sfxt,Bozzo2015:underluminous}. 
Viable models involve extremely dense inhomogeneities (``clumps'') in the winds 
of the SFXT supergiant companions, compared to classical systems \citep[][]{zand2005,Negueruela2008},
the presence of magnetic/cen\-trif\-u\-gal gates generated by the slower rotational 
velocities and higher magnetic fields of the neutron stars hosted in SFXTs 
\citep{Grebenev2007,Bozzo2008}, 
or  a subsonic settling accretion regime combined with magnetic reconnections between 
the NS and the supergiant field transported by its wind 
\citep{Shakura2012:quasi_spherical,Shakura2014:bright_flares}.

The \sw\ \citep[][]{Gehrels2004} 10 year anniversary also marks the completion of the 
first 7 years of the \emph{Swift}  SFXT Project, which has been investigating the 
properties of SFXTs with a strategy that exploits \sw's uniqueness. 
In this Paper I will review the main results of its long term monitoring and outburst 
follow-ups of an ever increasing sample of SFXTs, 
a monitoring that has recently extended to include a small control sample of classical systems. 
In particular, I will show the new results on the two sources IGR~J16493$-$4348 and 
AX~J1845.0$-$0433 which were observed during 2014, and put them in the broader 
context of the comparison of SFXT vs.\ classical systems.  
I will also show that, as customary, the seventh year harbors  
a ``seventh year crisis'', since the most recent observations seem to challenge the 
prevailing models that attempt to explain the SFXT behaviour.

            \section{Results from the \emph{Swift} SFXT project \label{swift10:results}}
 
A continuing effort to improve and fine-tune \sw's GRB observing strategy 
has allowed a gradual shift of overall observing time from mostly GRB science 
to (currently) mainly guest observer and target of opportunity (ToO) targets.  
Several initiatives could then be carried out that would boost \sw's secondary science 
by selecting well-defined astrophysical problems that could be 
effectively tackled by exploiting \sw's fast automatic slewing 
and  multi-wavelength capability, 
as well as its flexible observing schedule and very low overheads. 
The \sw\ SFXT Project was born as one of these initiatives. 
\begin{landscape}
\setcounter{table}{0}   
\begin{table*}[H] 
\tiny
 \tabcolsep 2pt         
 \begin{center}
 \caption{Summary of \sw/XRT campaigns divided by monitoring samples ({\it yearly} and {\it orbital}; Sect.~\ref{swift10:sec:longterm} and \ref{swift10:sec:monit2}). 
N is the number of observations (individual ObsIDs) obtained during the monitoring campaigns; N$_{\rm c}$ the number of observations used (Col.~6).  
Count rates (Col.~8) are in units of $10^{-3}$ counts s$^{-1}$ in the 0.2--10\,keV energy band, while   
observed fluxes (Col.~9) are in units of $10^{-12}$ erg cm$^{-2}$ s$^{-1}$ 
and luminosities (Col.~10)  in units of $10^{34}$ erg s$^{-1}$, both in the 2--10\,keV energy band; all are 
based on a single 900\,s exposure. 
    $\Delta T_{\Sigma}$ (Col.~11) is the sum of the exposures accumulated in all observations, 
   each in excess of 900\,s, where only a 3-$\sigma$ upper limit was achieved;  
   $P_{\rm short}$  (Col.~12) is the percentage of time lost to short observations; 
   IDC  (Col.~13) is the  {\it inactivity duty cycle}, 
   the time each source spends  undetected down to a flux limit of reported in Col.~9;  
   Rate$_{\Delta T_{\Sigma}}$ (Col.~14)  
   is the observed count rate in the data for which no detections were obtained as single observations (see Sect.~\ref{swift10:sec:dc}). }
 \label{swift10:tab:tab1}
 \begin{tabular}{rcc rrrr rrlrrr cc cc c }
 \hline
 \hline
 \noalign{\smallskip}
Name &  Period & Distance & Start & End &N$_{\rm c}/$N & Expo. &  CR$_{\rm lim}^{\rm 0.2-10}$  & $F_{\rm lim}^{\rm 2-10}$ & $L _{\rm lim}^{\rm 2-10}$&$\Delta T_{\Sigma}$  & $P_{\rm short}$ &  IDC$^{a}$ 
                  & Rate$_{\Delta T_{\Sigma}}^{0.2-10}$   & Ref$^{b}$  & Ref$^{c}$  & Ref$^{d}$ \\
  \noalign{\smallskip}
&  (d)&  (kpc)  & UT & UT  & & (ks) & ($\times10^{-3}$) &  ($\times10^{-12}$)  &($\times10^{34}$)
                &(ks) & (\%) &  (\%) &   ($\times10^{-3}$)    & P   &    D   &      \\  
  \noalign{\smallskip}
 (1)&  (2)&  (3)  & (4) & (5)  & (6)  & (7) & (8) &  (9)  &(10)   &(11) & (12) &  (13) &   (14)    & (15)   &   (16)   &  (17)      \\  
  \noalign{\smallskip}
 \hline
 \noalign{\smallskip} 
{\it Yearly sample} \\
IGR~J08408$-$4503 &  --       &3.4$\pm$0.35 &2011-10-20    &2012-08-05 &    77/82  &74.4   &17  &1.9   &0.26 &46.6   &7    &67.2$_{-5.7}^{+4.9}$  &7.2$\pm$0.6     & -- &12    &  18 \\   
IGR~J16328$-$4726 &10.076  &6.5$\pm$3.5   &2011-10-20    &2013-10-24 &    94/98  &88.0   &14  &2.7   &2.5   &47.5   &12  &61.0$_{-5.2}^{+4.8}$  &4.0$\pm$0.4     &1  &13   & 18  \\   
IGR~J16465$-$4507 &30.243  &12.7$\pm$1.3 &2013-01-20    &2013-09-01 &    61/65  &58.6   &16  &2.0   &4.4   &3.0     &0    &5.1$_{-1.6}^{+4.4}$    &14.6$\pm$0.4   &2  &12     & 18 \\   
IGR~J16479$-$4514 &3.3193  &4.9                   &2007-10-26    &2009-10-25 & 139/144 &159.8 &16  &2.5   &1.1   &29.7   &3    &19.4$_{-2.9}^{+3.8}$  &3.1$\pm$0.5     &3  &14   & 19,20,18 \\   
IGR~J16493$-$4348 &6.782   & $>6$               &2014-01-19    &2014-09-03 &    55/65  &52.7    &13  &2.2   &1.3    &5.8     &3    &11.3 $_{-3.0}^{+5.7}$  & $>7.9^{e}$     &4 & 15 & 21\\ %
XTE~J1739$-$302    &51.47    &2.7                   &2007-10-27     &2009-11-01 & 181/184 &206.6 &13  &1.6  &0.18  &71.5   &10  &38.8$_{-3.5}^{+3.8}$  &4.0$\pm$0.3     &5   &14  &  19,20,18 \\  
IGR~J17544$-$2619 &4.926   &3.6                    &2007-10-28    &2009-11-03 & 138/142 &142.5 &12  &1.4  &0.21  &69.3   &10  &54.5$_{-4.3}^{+4.1}$  &2.2$\pm$0.2     & 6   &14 & 19,20,18  \\ 
AX~J1841.0$-$0536 &6.4530 &7.8$\pm0.74$  &2007-10-26    &2008-11-15 &   87/88   &96.5   &13  &1.8   &1.6    &26.6  &3    &28.4$_{-4.3}^{+5.1}$  &2.4$\pm$0.4      & 7  & 12 & 19,18  \\ 
AX~J1845.0$-$0433 &5.7195 &  6.4$\pm$0.76  &2014-02-14    &2014-10-10 &   71/80   &69.1   &13  &1.9  &0.11  &7.9     &4    &11.8$_{-2.8}^{+4.9}$  &  3.5$\pm$0.9  &8 & 12 & 21\\  
  \noalign{\smallskip}
{\it Orbital sample} \\
IGR~J16418$-$4532  &3.73886 & 13                    &2011-02-18 &2011-07-30  & 15/15   &43.3 &19  &12.5   &36     &4.8     &0  &11.0$_{-3.8}^{+13.1}$  & $>9.2^{e}$     &9 &14 & 22,18 \\  
IGR~J17354$-$3255  &8.448    & 8.5                    &2012-07-18 &2012-07-28  & 22/22   &23.7 &14 &2.2      &3.3    &7.8     &1  &33.4$_{-8.3}^{+11.1}$  & $>4.6^{e}$     &10  &16 & 23,18\\  
IGR~J18483$-$0311  &18.545  &2.83$\pm$0.05 &2009-06-11 & 2009-07-08 & 23/23   &44.1 &11  &1.8     &0.24  &11.8   &0  &26.6$_{-7.1}^{+10.9}$  & $3.6\pm0.8$  &11  &17 & 24,18 \\  
  \noalign{\smallskip}
  \hline
  \end{tabular}
  \end{center}
\begin{list}{}{} 
\item[$^{\mathrm{a}}$]{Uncertainties obtained from or with the method described in \citet[][]{Romano2014:sfxts_paperXI}.}

\item[$^{\mathrm{b}}$]{References to orbital periods: 
(1) \citet{Corbet2010:16328-4726}; 
(2) \citet{LaParola2010:16465-4507_period}; 
(3) \citet{Romano2009:sfxts_paperV};
(4) \citet[][]{Cusumano2010:16493-4348_period}; 
(5)  \citet[][]{Drave2010:17391_3021_period};  
(6) \citet{Clark2009:17544-2619period}; 
(7) \citet[][]{Gonzalez2015:PhD}  
(8) \citet[][]{Goossens2013:18450_period}; 
(9) \citet{Levine2011}; 
(10)  \citet{Dai2011:period_17354}; 
(11)  \citet{Levine2006:igr18483}.}

\item[$^{\mathrm{c}}$]{References to distances: 
 (12) \citet[][]{Coleiro2013:distribHMXBs};  
 (13) \citet[][]{Fiocchi2012:16328-4726}; 
 (14) \citet{Rahoui2008}; 
 (15) \citet[][]{Nespoli2008:16493}; 
(16) \citet[][]{Tomsick2009:cxc17354}; 
(17) \citet{Torrejon2010:hmxbs}. }

\item[$^{\mathrm{d}}$]{References to original data papers: 
(18) \citet[][]{Romano2014:sfxts_paperX}; 
(19) \citet[][]{Romano2009:sfxts_paperV}; 
(20) \citet[][]{Romano2011:sfxts_paperVI}; 
 (21) This work; 
 (22) \citet[][]{Romano2012:sfxts_16418}; 
 (23) \citet[][]{Ducci2013:sfxts_17354}; 
 (24) \citet[][]{Romano2010:sfxts_18483}. 
}
\item[$^{\mathrm{e}}$]{3-$\sigma$ upper limit. }
\end{list}
  \end{table*} 
\end{landscape}

           \subsection{The long outburst of IGR~J11215$-$5952\label{swift10:sec:11215}} 

\begin{figure*}[t]
\vspace{-4.truecm}
\centerline{\includegraphics[width=16cm,angle=0]{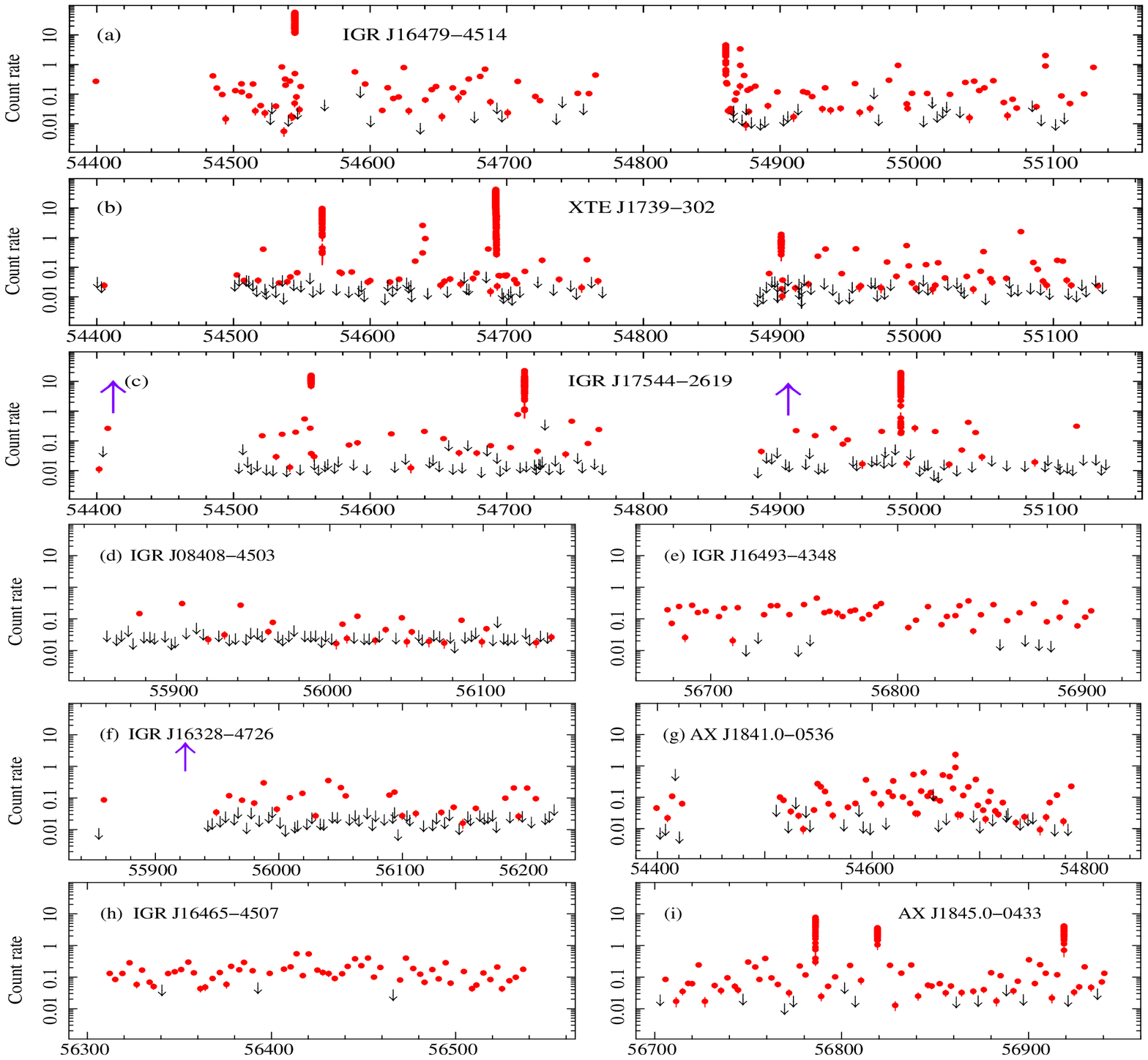}}
\vspace{-5.5truecm}
\caption[XRT campaigns]{{\it Swift}/XRT (0.2--10\,keV) 
long term  light curves for the {\it yearly monitoring sample}. The x-axis is in MJD. 
The points refer to the average count rate measured for each observation  except for outbursts,  
where the data were binned to include at least 20 counts bin$^{-1}$ (to best represent the 
dynamic range). Red points are detections, black downward-pointing arrows are 
3-$\sigma$ upper limits. 
Violet upward-pointing arrows mark bright outbursts detected by BAT (and not simultaneously followed by XRT),
or by MAXI (for AX~J1841.0$-$0536).  
The data on IGR~J16493$-$ 4348 (panel e) and 
AX~J1845.0$-$0433 (panel i) are presented here for the first time. 
The remainder of the data are adapted from papers listed in Table~\ref{swift10:tab:tab1} (Col.~17). 
}
\label{swift10:fig:campaigns}
\end{figure*}

The first experiment within the \sw\ SFXT Project was performed in February 2007 
\citep[][]{Romano2007} when we took advantage of the fact that at least one
SFXT, IGR~J11215$-$5952 (hereon J11215), showed outbursts with a periodicity of about 330\,d
\citep[][]{SidoliPM2006} in the  \inte\ data,
and monitored one outburst with the 
\sw/X-ray Telescope \citep[XRT, 0.2--10\,keV,][]{Burrows2005:XRT} 
from before onset, to the peak, and until 
it became undetectable with reasonable XRT exposures. 
The campaign lasted 23 d (total exposure of $\sim73$ ks) and 
showed that, differently from previously thought based on lower-sensitivity 
instruments observing only the brightest hr-long flares (e.g.\ the hard X-ray monitor on board \inte),
the soft X-ray emission, hence the accretion phase, lasted several days. 
Superimposed on the light curve, frequent flares are also observed, 
probably due to inhomogeneous accretion, i.e.\ clumps in the accreting wind. 
This phenomenology is consistent with a 
gradually increasing flux at the periastron passage in a wide eccentric orbit. 

\sw\ also allowed the determination of the true orbital period 
$P_{\rm orb}\sim164.6$\,d \citep[][]{Sidoli2007,Romano2009:11215_2008},  
a rare instance, as orbital periods are generally found from all-sky monitor data.

Furthermore, since the outbursts of this source are periodic and regularly spaced, 
and a non negligible eccentricity is required to account for the observed low 
quiescent luminosity, these data led \citet[][]{Sidoli2007} to 
propose the existence of a second wind component  
from the supergiant companion in addition to its normal symmetric polar wind.
This equatorially enhanced wind component is probably clumpy,  
denser and slower than the polar one,  
and inclined with respect to the orbital plane (to account for the narrowness 
of the outburst X-ray light curve).

             \subsection{Long term properties from monitoring campaigns\label{swift10:sec:longterm}} 

In the wake of the success of the J11215 campaign, 
during the Fall 2007 we selected 4 SFXTs from the 8 known at the time, 
IGR~J16479$-$4514, XTE~J1739--302, IGR~J17544$-$2619, and AX~J1841.0$-$0536 
(hereon J16479, J1739, J17544, and J1841, respectively, the seed of our 
 {\it yearly sample}, see Table~\ref{swift10:tab:tab1}), 
and set to perform a systematic study \citep[][]{Sidoli2008:sfxts_paperI} 
with a year-long series of 1--2\,ks pointed observations with XRT.  
At the time, the SFXT binary periods were largely unknown and believed to be in excess 
of 10\,d (see the current status in Table~\ref{swift10:tab:tab1}, Col.\ 2), 
so the observations were scheduled 3--4 days apart. 
The initial goals were to 
{\it i)} seek for the signatures of equatorial winds, 
{\it ii)} catch outbursts to determine whether they showed any periodicities, 
{\it iii)} follow the sources during the whole outburst duration, and 
{\it iv)} monitor the quiescence. 

For this project, the \sw/Burst Alert Telescope \citep[BAT, 15--150\,keV,][]{Barthelmy2005:BAT} Team 
applied the ``BAT special functions'' to the sample of SFXTs and candidates 
(i.e., sources with similar X-ray flaring behaviour but with no firm measurement 
of the spectral type of the companion). This allowed 
\sw\ to react to an increase in flux from any SFXTs as if they were a GRB.
In this way, whenever an SFXT triggers the BAT, simultaneous broad band data are collected that
span $\sim 1600$--6000\,\AA\ through the UV/Optical Telescope \citep[UVOT, ][]{Roming2005:UVOT},  
and 0.2--150\,keV through XRT and BAT combined. 

Before we began our investigation, deep \xmm\ exposures 
\citep[][]{Gonzalez2004} reporting fluxes between $\sim 10^{-13}$ and $10^{-10}$ erg s$^{-1}$
had described the characteristics of J17544 away from the bright outbursts, including 
a trend for harder spectra at higher fluxes, 
while a revealing \chandra\ observation \citep[][]{zand2005} 
had caught the first detection of a fast X-ray transient in quiescence, 
a state characterized by a very soft (photon index $\Gamma=5.0\pm1.2$) spectrum.
On the other hand, the long-term behaviour of SFXTs, not unlike any other hard 
X-ray transient, had been traditionally investigated only with coded-mask 
large field-of-view instruments, 
such as \inte/IBIS \citep[][]{Ubertini2003} or \sw/BAT which, because of their 
sensitivity limits, mostly catch only the brightest portion of any transient event 
\citep[see][for a catalogue of more than a thousand \sw/BAT 
bright SFXT flares described in Sect.~\ref{swift10:sec:N_SFXTs}]{Romano2014:sfxts_catI}.
Therefore, our strategy, by combining sensitive soft X-ray monitoring 
with outburst follow-ups, has allowed us through the years for the first time, 
to systematically assess the soft X-ray long term properties 
of a conspicuous fraction of the SFXT sample when away from the prominent 
bright outbursts, in particular, the states leading down to quiescence.

The detailed results on the first campaigns can be found in 
\citet[][]{Sidoli2008:sfxts_paperI} and \citet[][]{Romano2009:sfxts_paperV,Romano2011:sfxts_paperVI},  
those on 3 more sources, IGR J08408$-$4503, IGR J16328 $-$4726, 
and IGR J16465$-$4507 (hereon J08408, J16328, and J16465)  
in \citet[][]{Romano2014:sfxts_paperX}. 
Table~\ref{swift10:tab:tab1} reports the characteristics of the sources in our 
SFXT sample (binary period and distance, Cols.\ 2 and 3) and 
the campaign dates (Cols.\ 4 and 5).
In the following, we summarize the results for these sources and introduce the new ones 
for J16493  
and J1845,  
that were observed in 2014\footnote{The new data were processed 
\citep[see, e.g.,][]{Romano2014:sfxts_paperX} with standard software ({\sc FTOOLS} v6.16),  
calibration (CALDB 20140709), and methods ({\sc xrtpipeline}, v0.13.1).}. 

Fig.~\ref{swift10:fig:campaigns} shows the XRT light curves of the {\it yearly sample} 
at a daily resolution. The following can be observed: 
\begin{enumerate}
\item  
The most striking features are, of course, the bright outbursts, which will be described
in detail in Sect. \ref{swift10:sec:outburst}. 
These outbursts, surprisingly,  are however merely the tip of the iceberg in the SFXT activity
(only a few percent of the total time). 
\item  
All sources display a dynamic range (DR) of $\sim3$--4 orders of magnitude\footnote{We 
note that both J1841 which went into outburst after the end of the campaign 
\citep[2010 Jun 5,][]{Romano2011:sfxts_paperVII} 
and J1845 which went into outburst on 2012 May 05 \citep[][]{Romano2013:Cospar12}, 
reached  DR$\ge 10^{3}$. }.   
\item  
The exceptions to the previous points are J16465 and the newly observed 
J16493, the latter showing a DR$\sim48$. 
These two sources (Fig.~\ref{swift10:fig:campaigns}h and 
\ref{swift10:fig:campaigns}e, respectively), 
based on their small DR, are therefore found 
to be not SFXTs, but classical systems. 
\item  
For SFXTs, most of the emission is found outside the bright outbursts, 
so that the long-term behaviour of SFXTs is not quiescence 
but an intermediate state with an average X-ray luminosity of 
$10^{33}$--$10^{34}$ erg\,s$^{-1}$ 
(and, as reported below, a spectrum that can be fit with a power law with photon index 
$\Gamma=1$--2). 
\item  
Variability is observed at all timescales we can probe. 
Superimposed on the day-to-day variability, we measure intra-day flaring that 
involves flux variations up to one order of magnitude;  
we identify flares down to a count rate in the order of 0.1\,counts s$^{-1}$ 
($L\sim2$--$6\times10^{34}$ erg s$^{-1}$) within a snapshot of about 1\,ks. 
\end{enumerate} 

As shown by \citet[][]{Walter2007}  
the short time scale variability cannot be accounted for by accretion from a 
homogeneous wind, but it can naturally  be explained 
by the accretion of single clumps in the donor wind.
We calculated that the average clump mass is $M_{\rm cl} \sim 0.3$--$2\times10^{19}$ g  
\citep[][]{Romano2011:sfxts_paperVI}, 
about those expected \citep[][]{Walter2007} to be responsible for short flares, below the \inte{} 
detection threshold and which, if frequent enough, may significantly contribute to the 
mass-loss rate.

\begin{figure}[t]
\begin{flushleft}
\vspace{+0.3truecm}
\hspace{-0.3truecm}
\centerline{\includegraphics[height=9cm,angle=270]{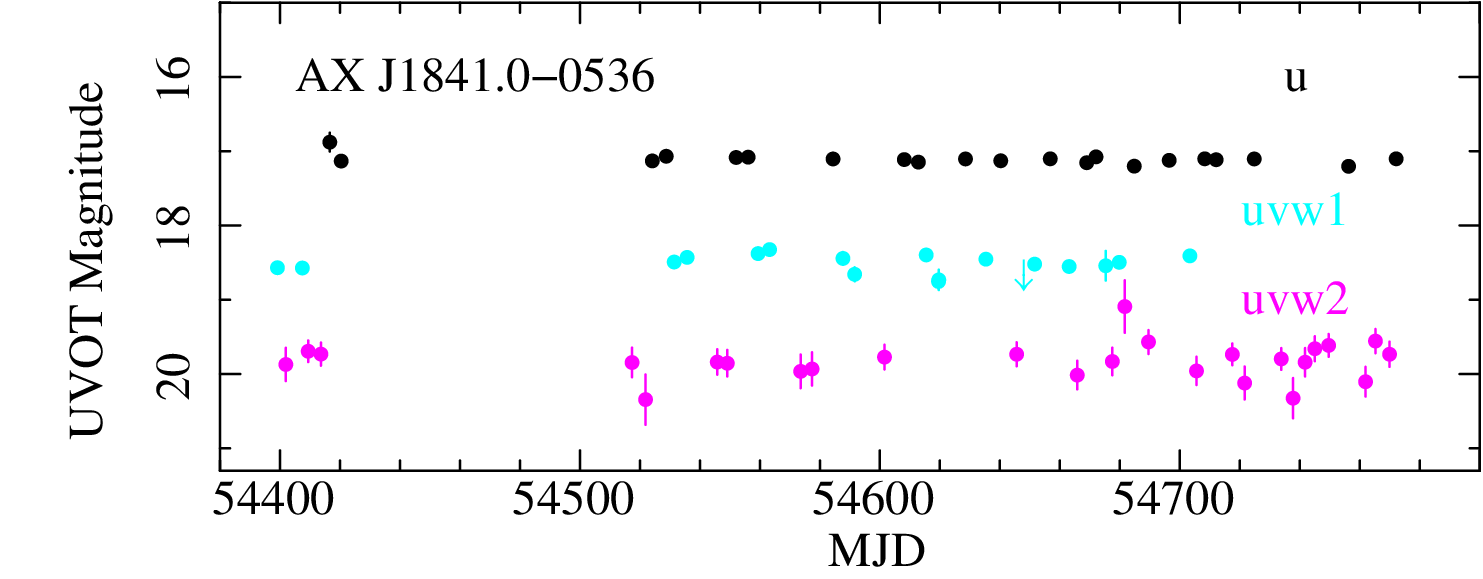}}
\end{flushleft}
\vspace{-0.5truecm}
\caption[XRT campaigns]{{\it Swift}/UVOT light curves of AX~J1841.0$-$0536. 
Adapted from \citet[][]{Romano2009:sfxts_paperV}.
}
\label{swift10:fig:campaigns2}
\end{figure}

The data collected during these campaigns were used to perform soft X-ray 
intensity-selected spectroscopy.  
We find that the common out-of-outburst spectroscopic properties are a 
non-thermal emission (power law with $\Gamma=1$--2) combined with 
a soft excess becoming increasingly more dominant as the source flux state 
becomes lower, and a ubiquitous harder-when-brighter trend 
\citep[][]{Romano2009:sfxts_paperV,Romano2011:sfxts_paperVI,Romano2014:sfxts_paperX}. 
Therefore, the spectral modelling of out-of-outburst emission shows that accretion 
is occurring down to very low luminosities. This is clearly at odds with what 
is generally observed in the hard X-rays, due to the different instrumental sensitivities involved.  
Since a non-thermal spectrum plus a soft excess is common in classical systems, 
the spectroscopic properties are not an efficient method for discriminating 
SFXTs within the HMXB sample, differently from what happens for the dynamic range. 

We note that when combining single 1--2\,ks snapshots in which no 
detection was individually achieved, the XRT data can reach luminosities comparable 
to the quiescent one \citep[][]{Romano2009:sfxts_paperV}. 
For instance, in the cases of J17544 and J1739 
\citep[][]{Romano2011:sfxts_paperVI} 
and J08408 \citep[][]{Romano2014:sfxts_paperX}, luminosities of a few 
$10^{32}$ erg\,s$^{-1}$, have been reached, and the spectral properties 
observed in this very low state are consistent with those observed with
deep \xmm\ exposures \citep[e.g.][]{Bozzo2010:quiesc1739n08408}. 

UVOT observed  simultaneously with the XRT during most of our monitoring 
with different combinations of optical and UV filters, depending on the magnitude of the 
companion stars \citep[][]{Romano2009:sfxts_paperV,Romano2011:sfxts_paperVI}. 
In J1739, only mar\-gin\-al variability was observed  in the $u$ and $uvw1$ filters; 
the $uvw1$ light curve of J17544 and the $u$ and $uvw1$ light curves of J1841 
were remarkably stable, consistently with the optical/UV emission being dominated by the 
constant contribution of the supergiant companions. Fig.~\ref{swift10:fig:campaigns2} shows, as an 
example, the UVOT light curves of J1841.

\begin{figure}[t]
\begin{flushleft}
\vspace{-0.4truecm}
\hspace{-0.5truecm}
\centerline{\includegraphics[width=9.5cm,angle=0]{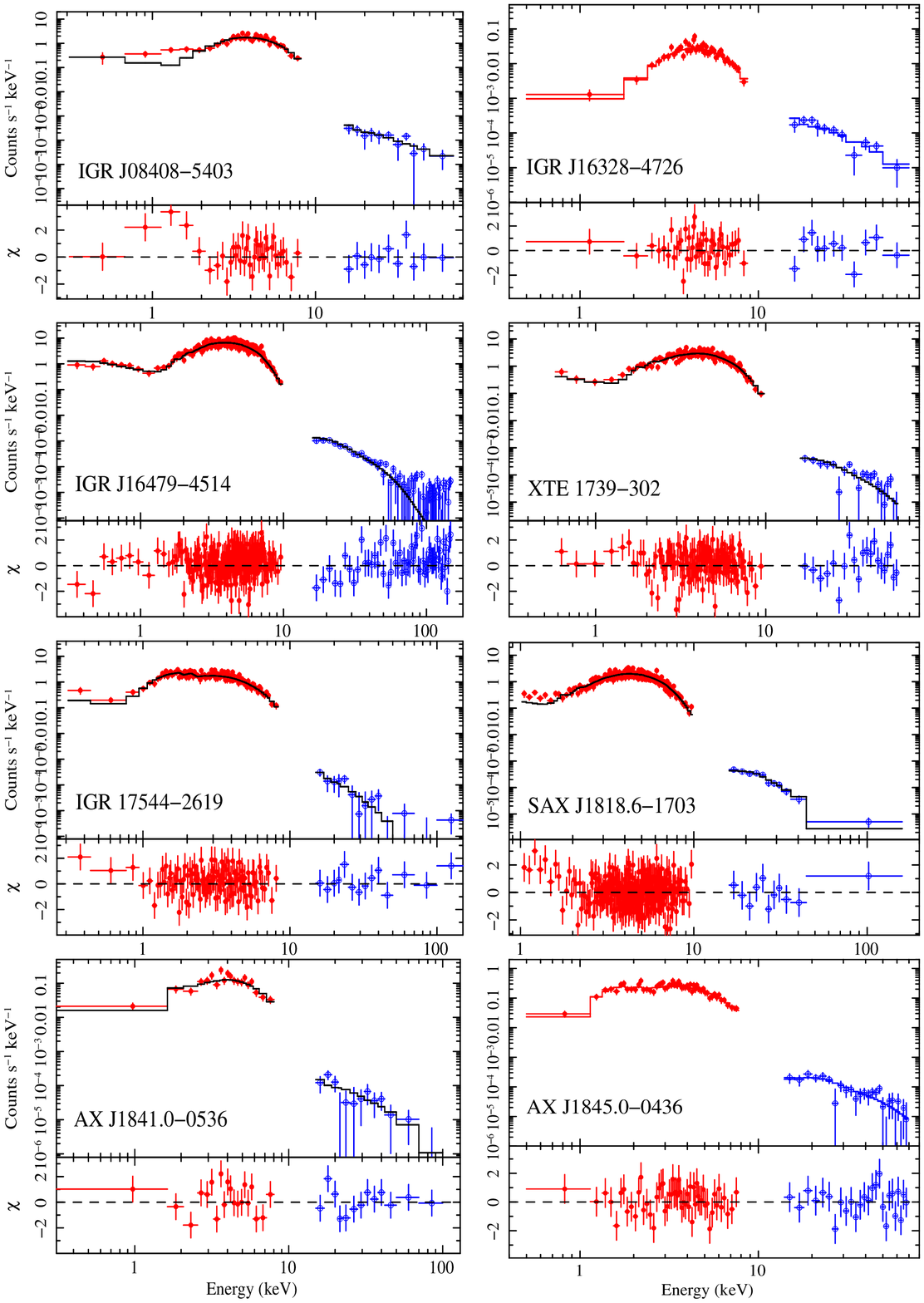}}
\end{flushleft}
\vspace{-1.2truecm}
\caption[XRT light curves]{ {\it Swift}/XRT and BAT simultaneous spectroscopy. 
Filled (red) circles and empty (blue) circles denote XRT and BAT data, respectively. 
The data are fit with an absorbed power law with a high energy cut-off  model or 
absorbed power laws with high energy exponential cut-off model.
Data references: 
IGR~J08408$-$4503   \citep[2008-07-05,][]{Romano2009:sfxts_paper08408}; 
IGR~J16328$-$4726   \citep[2009-06-10,][]{Romano2013:Cospar12}; 
IGR~J16479$-$4514   \citep[2008-03-19,][]{Romano2008:sfxts_paperII}; 
XTE~J1739$-$302      \citep[2008-08-13,][]{Sidoli2009:sfxts_paperIV};  
IGR~J17544$-$2619  \citep[2009-06-06,][]{Romano2009:sfxts_paperV}; 
SAX~J1818.6$-$1703 \citep[2009-05-06,][]{Sidoli2009:sfxts_sax1818};  
AX~J1841.0$-$0536   \citep[2010-06-05,][]{Romano2011:sfxts_paperVII}; 
AX~J1845.0$-$0433   \citep[2012-05-05,][]{Romano2013:Cospar12}.
 } 
\label{swift10:fig:broadspectra}
\end{figure}

%
\begin{figure}[t]
\begin{flushleft}
\vspace{-0.3truecm}
\centerline{
\hspace{+0.2truecm}
\includegraphics[width=9cm,angle=0]{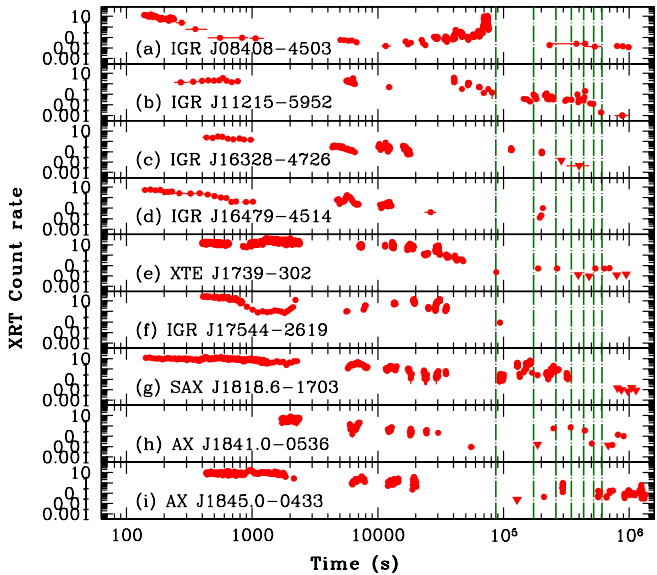}}
\end{flushleft}
\vspace{-2.0truecm}
\caption[XRT light curves]{ {\it Swift}/XRT light 
curves of the better followed-up outbursts of confirmed SFXTs, 
referred to their respective BAT triggers (except: IGR~J11215$-$5952 is   
referred to MJD 54139.94). 
Points mark detections, triangles 3$\sigma$ upper limits, 
vertical dashed lines mark daily intervals, up to a week. 
Data references: 
IGR~J08408$-$4503   \citep[2008-07-05,][]{Romano2009:sfxts_paper08408};
IGR~J11215$-$5952   \citep[2007-02-09,][]{Romano2007}; 
IGR~J16328$-$4726   \citep[2009-06-10,][]{Romano2013:Cospar12}; 
IGR~J16479$-$4514   \citep[2005-08-30,][]{Sidoli2008:sfxts_paperI}; 
XTE~J1739$-$302      \citep[2008-08-13,][]{Sidoli2009:sfxts_paperIV}; 
IGR~J17544$-$2619   \citep[2010-03-04,][]{Romano2011:sfxts_paperVII}; 
SAX~J1818.6$-$1703 \citep[2009-05-06,][]{Sidoli2009:sfxts_sax1818};  
AX~J1841.0$-$0536   \citep[2010-06-05,][]{Romano2011:sfxts_paperVII}; 
AX~J1845.0$-$0433   \citep[2012-05-05,][]{Romano2013:Cospar12}.
Adapted from \citet[][]{Romano2013:Cospar12}. 
}
\label{swift10:fig:lcvs_best_sfxts}
\end{figure}

           \subsection{Orbital monitoring campaigns\label{swift10:sec:monit2}} 

Further monitoring campaigns were also performed  on 
three more sources with higher-cadence pointed observations 
(several XRT snapshots a day to provide intra-day sampling) 
for one or more orbital periods with the 
main goal of studying the effects of orbital 
parameters on the observed flare distributions.
In 2009 we performed the very first complete monitoring of the soft X-ray activity 
along an entire orbital period ($P_{\rm orb}\sim 18.5$\,d) of an SFXT, IGR~J18483$-$0311 
\citep[J18483,][]{Romano2010:sfxts_18483}. 
Similarly, in 2011 we monitored the candidate SFXT IGR J16418$-$4532 
\citep[J16418,][]{Romano2012:sfxts_16418} which has a 
much shorter orbital period, $P_{\rm orb}\sim 3.74$\,d 
and, in 2012,  the (then candidate) SFXT IGR J17354$-$3255 (J17354), with 
$P_{\rm orb}=8.4474$\,d \citep[][]{Ducci2013:sfxts_17354}.  
These three sources compose our {\it orbital monitoring sample}. 

These unique datasets allowed us to constrain in these objects the different mechanisms 
proposed to explain their nature. 
In particular, we applied the clumpy wind model for blue supergiants 
\citep[][]{Ducci2009} to the observed X-ray light curve. 
By assuming for J18483 an eccentricity of $e = 0.4$ 
and for J16418 circular orbits, 
we could explain their X-ray emission in terms of the accretion from a spherically 
symmetric clumpy wind, composed of clumps with different masses, 
ranging from $10^{18}$ to $\times10^{21}$\,g  for J18483,
and from $\sim5\times10^{16}$~g to $10^{21}$~g for J16418. 
Since J18483 is an intermediate SFXT with a moderately high dynamic range in the 
X-ray luminosity, the estimated sizes and masses of the clumps in J18483 
are somewhat larger that what would be expected according to the 
multidimensional simulations of massive stars winds 
\citep{Dessart2002,Dessart2003,Dessart2005} but likely not 
unrealistic \citep[see, e.g.][]{Furst2014:velax1}. 
The addition of magnetic/cen\-trif\-u\-gal gates could lower the requirements 
on the clump sizes and masses, but such mechanisms cannot be readily 
applied to the case of J18483 as the magnetic field of the NS hosted in 
this source and its spin period are highly debated \citep[e.g.][]{Ducci2013:18483,Sguera2015:18483}.
We found  that  J17354 is probably a wind-fed system and that the 
dip observed in its light curve cannot be explained with a luminosity modulation due to a 
highly eccentric orbit; on the contrary, it can be explained  in terms of an eclipse or the
onset of gated mechanisms.

\begin{figure*}[t]
\begin{center}
\vspace{-4.0truecm}
\centerline{\includegraphics*[angle=0,width=18.5cm,height=23cm]{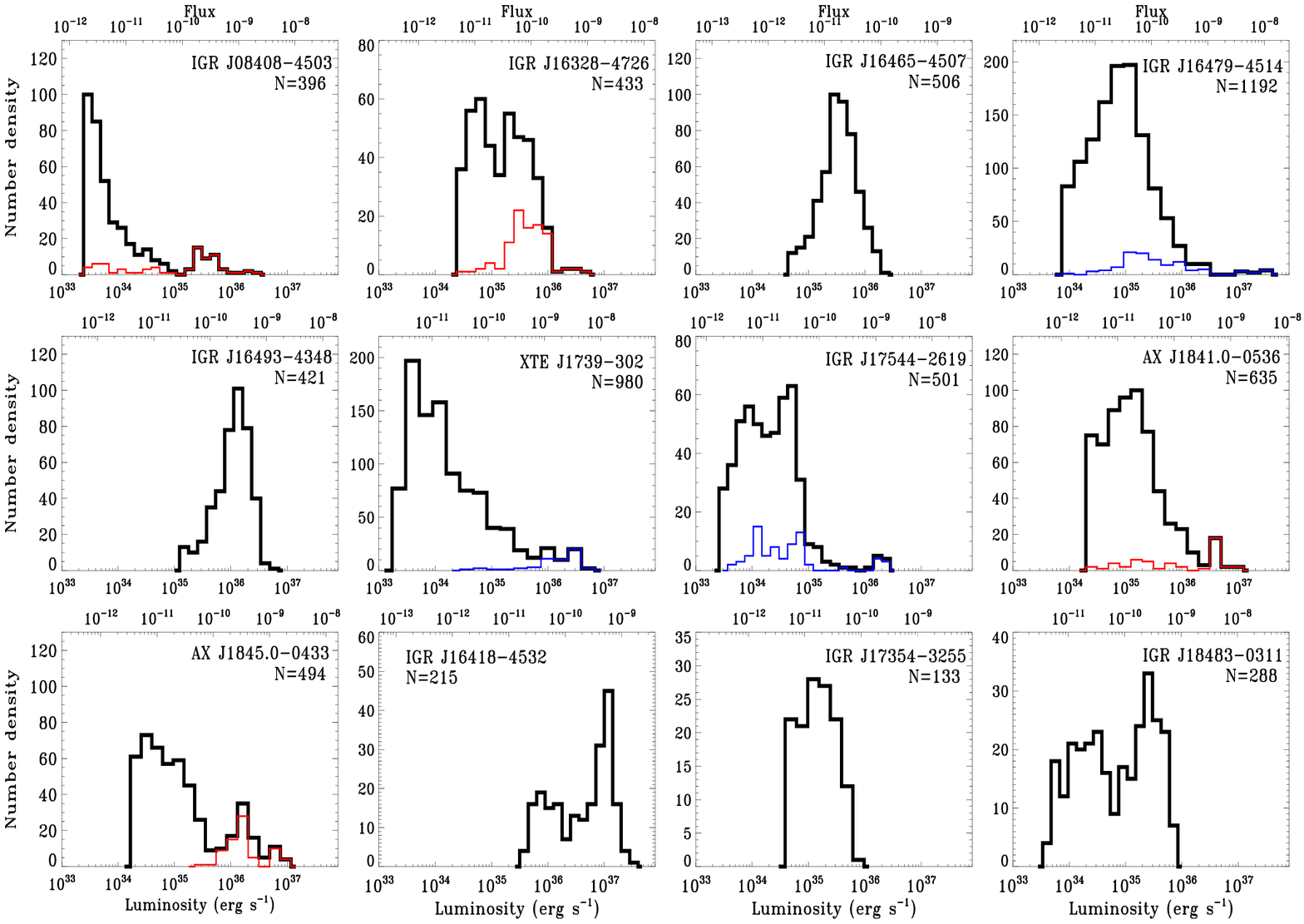}} 
\end{center}
\vspace{-8truecm}
\caption{ 
Differential distributions of the 2--10\,keV  luminosity (lower axis) and flux (unabsorbed, upper axis)  
drawn from the \sw/XRT light curves binned at 100\,s. N is the sample size. 
The thick black lines represent the data collected during
the monitoring campaigns (see Table~\ref{swift10:tab:tab1}, Col 4 and 5 for the campaign dates). 
The thin blue histograms show the part of the data collected as outburst observations during the campaign, 
thus including both the initial bright flare and the follow-up observations (see Sect.~\ref{swift10:sec:DLD}). 
The thin red histograms show outburst observations collected outside 
of the monitoring campaign (one outburst per source). 
The data on IGR~J16493$-$4348 and AX~J1845.0$-$0433 are presented here for the first time, 
the rest are adapted from \citet[][]{Romano2014:sfxts_paperX}.  
}
\label{swift10:fig:histos}
\end{figure*}

            \subsection{SFXT broad-band properties and arcsecond localizations 
              from outbursts and outburst follow-ups\label{swift10:sec:outburst}}  

Since \sw\ is a GRB-chasing mission, provided with 
fast automated slewing and panchromatic sensitivity, 
once SFXTs were included in the BAT special functions (see Sect.~\ref{swift10:sec:longterm}), 
SFXT outbursts started triggering the BAT and narrow field 
instrument (NFI) data (XRT and UVOT) started being collected
within a few hundred seconds (down to about $\sim 100$\,s) 
from the BAT trigger. 
The shape of the SFXT spectrum in outburst is 
a power law with an exponential cutoff at a few keV, therefore 
the large \sw{} energy range can both help constrain the 
hard-X spectral properties (to compare with popular 
accreting neutron star models) and measure the absorption. 

These simultaneous XRT and BAT data allowed us to perform, for the first time, 
simultaneous broad band spectroscopy of an SFXT in outburst 
\citep[J16479,][]{Romano2008:sfxts_paperII}. 
At the time of writing, we have collected a total of 51 bright flares (55 triggers, of which 4 double) 
that triggered the BAT, more than half of which followed-up with the NFI, 
thanks to the BAT special functions, so that 
we have been able to observe 
in this fashion most of the SFXT sample, as shown in Fig.~\ref{swift10:fig:broadspectra}. 
We found out that for all sources a good fit could indeed be obtained 
with a power-law with an exponential cutoff. 
This, in turn implies, based on the cutoff energy, 
a magnetic field consistent with a few $10^{12}$ G, 
typical of accreting NSs in HMXBs.  

The availability of such data also motivated the development of 
a physical model, {\tt compmag} in {\tt XSPEC} \citep[][]{Farinelli2012:compmag}, 
which includes thermal and bulk Comptonization for cylindrical accretion onto a 
magnetized neutron star. 
A full description of the algorithm \citep[see][]{Farinelli2012:compmag} 
is beyond the scope of this paper but the model  has been successfully applied to 
the SFXT class prototypes J17391 and J17544 
\citep[][]{Farinelli2012:sfxts_paperVIII},  
and to  J18483 \citep[][]{Ducci2013:18483}.  

An important benefit of NFI observations is \sw's ability to provide 
arcsecond localization for several SFXTs and candidates 
whose coordinates were only known to the arcminute level, 
or to improve on previously known coordinates
\citep[e.g.][]{Kennea2005:16479-4514,Grupe2009:16328-4726}. 
This greatly helps in associating with optical counterparts. 

Our observing strategy also includes XRT follow-ups 
for days (generally up to a week) after the outburst via ToO observations,  
well after it had become undetectable with monitoring instruments with lower 
sensitivity. 
Fig.~\ref{swift10:fig:lcvs_best_sfxts} shows the best examples of outburst light curves 
as observed by XRT, and exemplifies the common X-ray characteristics of this class:  
\begin{itemize}
\item  
extended soft X-ray activity around an outburst lasting up several days (see 
the vertical lines in Fig.~\ref{swift10:fig:lcvs_best_sfxts});  
\item  
a multiple-peaked structure;   
\item  
a DR (only including bright outbursts) up to $\sim3$ orders of magnitude. 
\end{itemize}

            \subsection{The 100-month SFXT BAT catalogue and the number 
of SFXTs in the Galaxy\label{swift10:sec:N_SFXTs}}

Since BAT observes an average of 88\% of the sky daily, it is ideally suited to 
detect flaring hard X-ray astrophysical sources, SFXTs in particular. 
We have thus produced the 100-month \sw\ Catalogue of SFXTs 
\citep[][]{Romano2014:sfxts_catI} which collects over a thousand BAT 
flares from 11 SFXTs, and reaches down to 15--150\,keV fluxes of about 
$6\times10^{-10}$ erg cm$^{-2}$ s$^{-1}$  (daily timescale) and about 
$1.5\times10^{-9}$ erg cm$^{-2}$ s$^{-1}$ (\sw\ orbital timescale).  
We found that these 
hard X-ray flares typically last at least a few hundred seconds, reach above 
100\,mCrab (15--50 keV), and last much less than a day.  
Their clustering in the binary orbital phase-space, however, demonstrates that these 
short flares are part of much longer outbursts, lasting up to a few days, 
as previously observed during our outburst follow-ups (Sect.~\ref{swift10:sec:outburst}). 
This large dataset can therefore probe the high and intermediate 
emission states in SFXTs, and help infer the properties of these binaries; 
it can also be used to estimate the number of flares per year each source is likely to 
produce as a function of the detection threshold and limiting flux in future missions. 
Finally, the catalogue has recently been exploited by \citet[][]{Ducci2014:sfxtN}  
to estimate  the expected number of SFXTs in the Milky Way, $N\approx 37{+53\atop-22}$. 
This shows that SFXTs constitute a large portion of X-ray 
binaries with supergiant companions  in the Galaxy.

            \section{Differential luminosity distributions \label{swift10:sec:DLD}}

An often understated property of the monitoring data is that  
{\it the yearly campaigns are statistically representative of the long-term soft X-ray properties of SFXTs} 
that the deep exposures from pointed telescopes can only rarely and non-uniformly sample. 
{\it Our observations are also independent}, since each observation is not triggered by the 
preceding ones (we consider the outburst followups as a separate set of data when in need 
of a statistical set). 
Our monitoring pace thus provides a casual sampling of the soft X-ray light curve
at a resolution of $\sim 3$--4\,d over a baseline of one or two years, 
therefore it offers both coverage of a large number of binary orbital cycles 
(ranging from $\sim 15$ cycles for  J17391  to $\sim 220$ for  J164794)
and a good sampling of the orbital phase \citep[see fig.~6 of][]{Romano2014:sfxts_paperX}.

Based on these premises, we can effectively calculate the percentage of time each 
source spent in different flux states, among which we distinguish:  
\begin{enumerate}
\item 
the flares that trigger the BAT (see Sect.~\ref{swift10:sec:outburst}) accounting for 
3--5\,\% of the exposure time;  
\item 
the intermediate states (all observations yielding a firm detection excluding outbursts); 
\item 
non detections (significance below 3$\sigma$; see Sect.~\ref{swift10:sec:dc}).   
Only observations with an exposure in excess of 900\,s were considered to account for  
non detections obtained during very short exposures (due to our observations being
interrupted by a higher figure-of-merit GRB). 
These correspond to flux limits $F_{\rm 2-10\,keV}^{\rm lim}\sim(1$--$3)\times 10^{-12}$\,erg\,cm$^{-2}$\,s$^{-1}$
(see Col.~9 of Table~\ref{swift10:tab:tab1}, and also the corresponding limits in count rates, Col.~8, 
and luminosity, Col.~10). 
\end{enumerate}

From the XRT light curves binned at 100\,s, of both the yearly and the orbital monitoring samples, 
and after removing the observations where a detection was not achieved (thus selecting only the intermediate states), 
we construct the 2--10\,keV differential luminosity distributions (DLD), shown in Fig.~\ref{swift10:fig:histos}
(solid black lines). 
For all sources we adopted a single conversion factor
between count rates, fluxes and luminosities, that were derived 
from the `medium'  spectrum for J16465, J16479, J16493, J1739, J17544, J1841,  J1845,  and J18483, 
the `low' spectrum for J08408 and J16328, the first observation for J16418, and the average spectrum
for the weak source J17354 \citep[][]{Romano2010:sfxts_18483,Romano2011:sfxts_paperVI,Romano2012:sfxts_16418,Ducci2013:sfxts_17354,Romano2014:sfxts_paperX}. 
Because the uncertainty in this conversion is dominated by those on the distance
(Table~\ref{swift10:tab:tab1}, Col.~3), the top x-axis of Fig.~\ref{swift10:fig:histos} also 
reports the flux scale in the same energy band. 
We note that the DLDs drawn from the orbital monitoring sample need to be taken with caution, as they
follow an entirely different observing strategy.
These observations were in fact collected with intensive campaigns during one or a few orbital periods,
as opposed to the few points per each period typical of the yearly campaign data, so 
short timescale variability may play an important role. 

Fig.~\ref{swift10:fig:histos} distinguishes (as a thin blue line) the 
data that were taken during an outburst that occurred during the observing 
campaign (two for J16479, and three for J1739 and  J17544), 
and those (thin red line) of outbursts that were observed outside of the
campaigns (we considered one for each of J08408, J16328, J1840, and J1845).

The DLD of the SFXT prototypes, 
J1739 and J17544, as well as J16479 and J08408 show two distinct populations of flares.
The first one is due to the outburst emission (peaking/reaching a few $10^{-9}$ erg cm$^{-2}$ s$^{-1}$), 
the second is due to the out-of-outburst emission, characterized by 
emission spanning up to 4 orders of magnitude in DR (at 100\,s binning). 
This also applies to the newly observed J18450, which also shares a DR  
of at least 3 orders of magnitude at a temporal resolution of 100\,s. 
We cannot exclude that particular 
distributions of the clump and wind parameters may produce a double-peaked DLD 
\citep[][]{Romano2014:sfxts_paperX},  
but this behaviour is more easily explained in terms of different accretion regimes as predicted 
by the magnetic/centrifugal gating model or the quasi-spherical settling accretion model 
\citep[][]{Grebenev2007,Bozzo2008,Shakura2012:quasi_spherical,Shakura2013:off_states}. 

The classical systems (J16465 and the newly observed J16493), on the contrary, only show 
one peak in their DCD, which is significantly brighter than those of of SFXTs.  
This confirms the findings of \citet[][]{Lutovinov2013:HMXBpop} that 
SFXTs show a median luminosity beneath the one of normal wind-fed HMXBs;  
the flaring observed in SFXTs can therefore be explained if some mechanism, such as 
magnetic arrest, can inhibit accretion.     
DCDs, therefore, can be used effectively to discriminate between the most extreme SFXTs, 
the intermediate systems,  
and classical systems. 

\begin{figure}
\begin{center}
\vspace{-0.2truecm}
\hspace{-0.1truecm}
\centerline{\includegraphics*[angle=90,width=9cm,height=6cm]{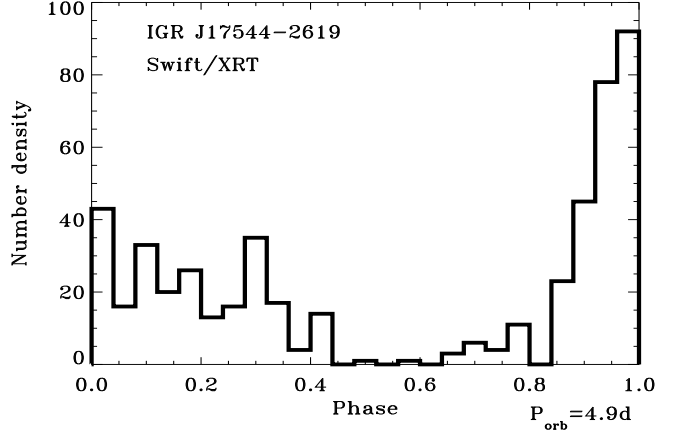}}
\end{center}
\vspace{-0.5truecm}
\caption{Distribution of the XRT detections (count rates) of  IGR~J17544$-$2619 
(0.2--10\,keV) folded at the orbital period, based 
on the most recent outburst ephemeris for this source 
\citep[$P=4.92693\pm0.00036$\,d, periastron at MJD $53732.65\pm0.23$;][]{Smith2014:atel6227}. 
}
\label{swift10:fig:17544phase}
\end{figure}

While the outbursts account for 3 and 5\,\% of the exposure time, the 
most probable flux level at which a random observation
will find these sources, when detected, can be retrieved from the peak of the DLD,
and this is of course a powerful tool to plan further observing campaigns, as is the 
distribution of detections along the orbital phase \citep[see fig.~6 of][]{Romano2014:sfxts_paperX}.
In particular, in Fig.~\ref{swift10:fig:17544phase} we show the case of J17544 
which flares tend to cluster around periastron more than in other SFXTs.

            \section{Inactivity duty cycles\label{swift10:sec:dc}}

The duty cycle of astrophysical sources is usually defined as the fraction of time 
the sources are active, and it is used to both characterize their emission properties and  
and to plan further observing campaigns to study them. 
SFXTs, however, show a very large dynamical range, with activity observed by the XRT
spanning several orders of magnitude in flux. It is more interesting, therefore, to define 
a measurement of inactivity as opposed of one of activity. 

From the non detections, we define the  {\it inactivity} duty cycle \citep[][]{Romano2009:sfxts_paperV}  
as the time each source spends {\it undetected} down to a flux limit of 
1--3$\times10^{-12}$ erg cm$^{-2}$ s$^{-1}$,  
\begin{equation}
{\rm IDC}= \Delta T_{\Sigma} / [\Delta T_{\rm tot} \, (1-P_{\rm short}) ] \, ,  
\end{equation}
where  
$\Delta T_{\Sigma}$ is the sum of the exposures (each longer than 900\,s) accumulated in all observations 
where only a 3$\sigma$ upper limit was achieved 
(Table~\ref{swift10:tab:tab1},  Col.\ 11), 
$\Delta T_{\rm tot}$ is the total exposure accumulated (Table~\ref{swift10:tab:tab1}, Col.\ 7), 
and $P_{\rm short}$ is the fraction of time lost to short observations  
(exposure $<900$\,s, Table~\ref{swift10:tab:tab1}, Col.\ 12). 
The cumulative count rate for each object is also reported 
Table~\ref{swift10:tab:tab1} (Col.\ 14).  

%
\begin{figure}
\begin{center}
\vspace{-0.3truecm}
\centerline{\includegraphics*[angle=0,width=9.5cm]{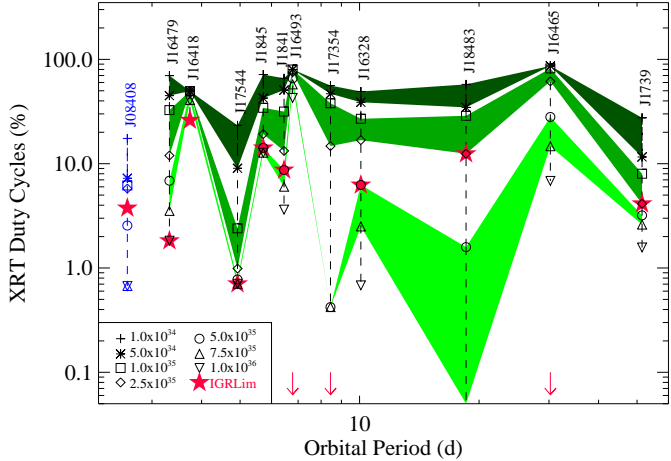}}
\end{center}
\vspace{-1truecm}
\caption{{
The XRT duty cycle (2--10\,keV) as a function of orbital period and for a range of 
2--10\,keV luminosities (see legend, in units of erg s$^{-1}$) in black.   
Only points above 0.1\,\% are shown. 
Since the period of  J08408 is currently unknown 
(see Table~\ref{swift10:tab:tab1} Col.~2 for the values adopted for the rest of the sample),
we arbitrarily place it at an orbital period of about two days and mark it in blue. 
The shaded areas mark the loci of the XRT duty cycle defined by contiguous luminosities.  
The red filled stars represent the XRT duty cycle at the \inte{} sensitivity for each object
(the downward-pointing arrows are consistent with 0).} 
The data on IGR~J16493$-$4348 and AX~J1845.0$-$0433 are presented here for the first time, 
the rest are adapted from \citet[][]{Romano2014:sfxts_paperX}.  
}
\label{swift10:fig:porb_dc}
\end{figure}

The need to provide uncertainties on IDCs by avoiding the standard approach of 
deriving them from extensive and time-consuming Monte Carlo bootstrap simulations 
has lead us to propose an application of Bayesian techniques, instead 
\citep[][]{Romano2014:sfxts_paperXI}. 
We exploited the fact that SFXTs are, when considering duty cycles, 
two-state sources, since they can only be found in either of two 
possible, mutually exclusive states, inactive (off) or flaring (on). 
Or, in this case, above or below a given flux threshold. 
We derived the theoretical expectation value for the duty cycle and its error 
as based on a finite set of independent observational data points 
following a Bayesian approach \citep[]{Romano2014:sfxts_paperXI}. 
The IDCs and their uncertainties thus calculated are reported in Table~\ref{swift10:tab:tab1}, Col.~13. 

We note that IDCs can be quite small for classical systems, which are to all extents and purposes 
persistent sources, and because they are on average more luminous. 
Once again, IDCs can help discriminate between SFXTs, intermediate and classical systems.

            \section{Discussion: the seventh year crisis\label{swift10:sec:discussion}}
 
In the following, we discuss the ``seventh year crisis'', 
the challenges that the recent observations (those collected by \sw, {\it in primis}) 
are making to the prevailing models attempting to explain the SFXT behaviour.

 {\bf Duty Cycles and orbital geometry. }      
If the properties of the binary geometry and inhomogeneities 
of the stellar wind from the primary were the leading causes of the observed 
X-ray variability in SFXTs, as initially proposed in clumpy wind models 
\citep[e.g.][]{zand2005,Negueruela2008,Walter2007}, 
generally larger IDCs would be expected for longer orbital periods.  
Naturally,  the definition of duty cycle is strongly dependent on the luminosity 
assumed as lower limit for the calculation. 
Therefore,  we exploited the high sensitivity afforded by the XRT observations and 
defined an {\it XRT luminosity-based duty cycle} (XRTDC) as the percentage of time
the source spends above a given luminosity. 
We considered several (2--10\,keV) luminosities in the range 
$L_{\rm 2-10\,keV}=10^{34}$--$10^{36}$ erg s$^{-1}$ 
and included the particular value of the luminosity corresponding to the \inte{} sensitivity 
for each object. 

Figure~\ref{swift10:fig:porb_dc}  
shows the XRTDC  as a function of the orbital period
with red stars marking the value at the \inte{} sensitivity.  
No clear correlation is found between the orbital periods and any of the duty cycles. 
This implies that wide orbits are not characterized by low duty cycles, 
as the clumpy wind models would predict. 
An intrinsic mechanism instead seems to be more likely responsible 
for the observed variability in SFXTs, i.e., either the 
wind properties or the compact object properties. 
However, it is hard to justify the radically different wind properties in SFXTs 
from those in classical systems with the same companion spectral type. 
Therefore, in light of this lack of correlation with the orbital period and 
our finding distinct flare populations (see Sect.~\ref{swift10:sec:DLD}),  
it seems more plausible that accretion-inhibition mechanisms 
or a quasi-spherical settling accretion regime may be in action, instead.

%
\begin{figure*}
\begin{center}
 
\vspace{+0.truecm}
\centerline{\includegraphics*[angle=90,height=7.9cm]{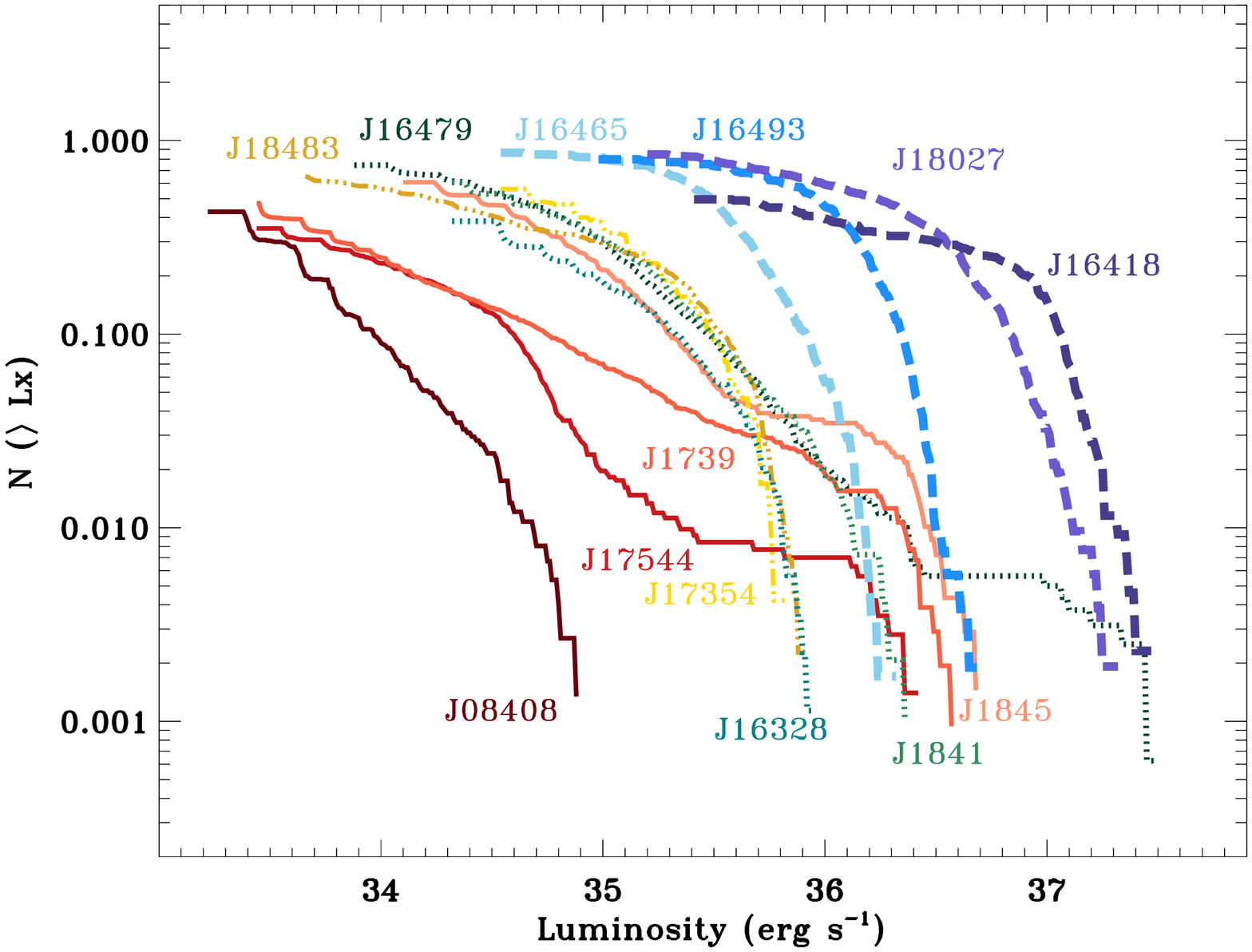}
\hspace{-2.3truecm}
\includegraphics*[angle=90,height=7.9cm]{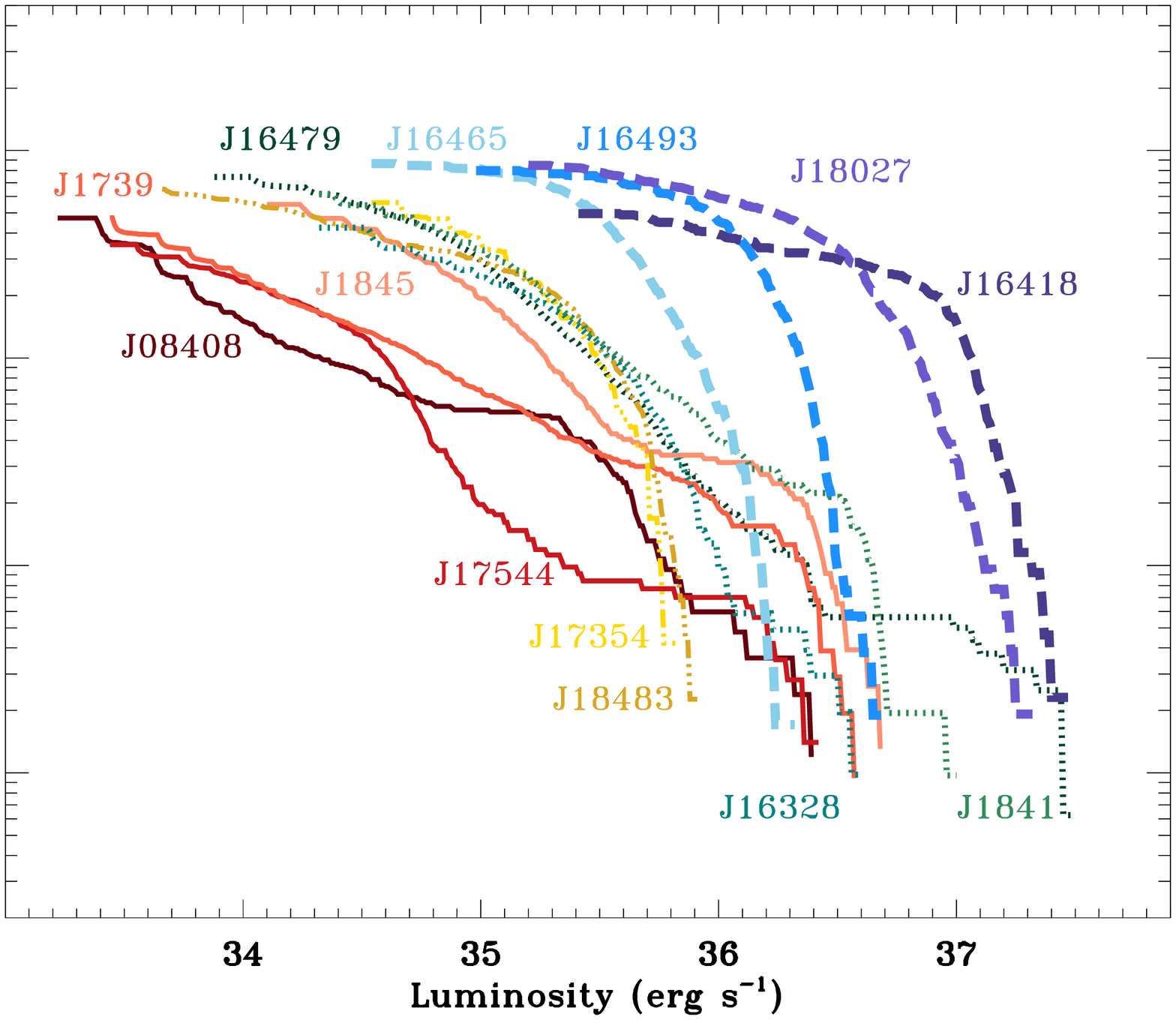}}

\end{center}
\vspace{-0.7truecm}
\caption{
Cumulative luminosity distributions of the 2--10\,keV  luminosity drawn 
from the \sw/XRT light curves binned at 100\,s. 
The classical SGXB  IGR~18027$-$2016 is marked with a thick dashed line.   
Newly classified classical systems IGR~J16465$-$4507, IGR~J16418$-$4532, and IGR~J16493$-$4348
are also marked with thick dashed lines. 
The intermediate SFXTs  IGR~J18483$-$0311 and  IGR~J17354$-$3255 
are shown as dot-dashed lines. 
IGR~J16328$-$4726, IGR~J16479$-$4514, and AX~J1841.0$-$0536 are shown as dotted lines,
while the  most extreme SFXTs 
   (IGR~J08408$-$4503, XTE~J1739$-$302, IGR~J17544$-$2619, and AX~J1845.0$-$0433) 
   are marked with solid lines.
The data on IGR~J16493$-$4348 and AX~J1845.0$-$0433 are presented here for the first time, 
the rest are adapted from \citet[][]{Bozzo2015:underluminous}.  
{\bf Left: } 
Only XRT data collected during the monitoring campaigns of all sources are used. 
{\bf Right: }  
Same as for the left, but in this case we also considered 
one outburst for the sources IGR~J08408$-$4503,  IGR~J16328$-$4726, AX~J1841.0$-$0536, and AX~J1845.0$-$0433  
recorded by the XRT outside the corresponding monitoring campaigns. 
}
\label{swift10:fig:CLDs}
\end{figure*}

{\bf Cumulative luminosity distributions. }   
In Sect.~\ref{swift10:sec:DLD} and \ref{swift10:sec:dc}  we have shown that 
DLDs and IDCs can be used effectively to discriminate SFXTs from classical systems, 
since the former are characterized by lower average luminosities. 
Another way to examine this property is to use the cumulative luminosity distributions (CLD) 
for our sample as calculated from the long term monitoring data (Sect.~\ref{swift10:sec:DLD}).
In \citet[][]{Bozzo2015:underluminous} the CLDs of the SFXT sample, as well as 
that of the classical SGXB IGR J18027 $-$2016 (J18027), are reported in the soft X-rays. 
Previous work constructed CLDs based on the \rxte{} Galactic bulge scan programme data \citep[][]{Smith2012:FXRT_RXTE} 
and \inte{} long-term monitoring \citep[][]{Paizis2014}.
In Fig.~\ref{swift10:fig:CLDs}, in which we also added the newly observed J16493 and J1845, 
the CLDs are normalized to the total exposure for each source, so that  the source duty cycle 
corresponds to the highest value on the y-axis. 
We can see that classical systems are characterized by CLDs with a 
single knee at $\sim10^{36}$--$10^{37}$ erg s$^{-1}$. 
On the contrary, SFXTs are systematically sub-luminous, with  their CLDs 
shifted at 100--100 times lower luminosities. 

As shown in Fig.~\ref{swift10:fig:CLDs}, 
the classical SgXRB  J18027 (thick dashed line) is characterized by a single knee at 
$\sim10^{36}$ erg s$^{-1}$.   
The CLDs of J16465, J16418, and J16493 (thick dashed lines)
closely resemble that of classical systems, with a position of the knee that can be accounted for 
once the relative distances and dependence of the flux from the orbital periods are considered.
As done in \citet[][]{Bozzo2015:underluminous} for the former two, we here reclassify the newly observed
SFXT candidate J16493 as a classical system. 
The intermediate SFXTs J18483 and  J17354 (dot-dashed lines) 
have CLDs similar to those of classical systems, but shifted at ten times lower luminosities.  
The CLDs of  J16328, J16479, and J1841.0 (dotted lines),   
are characterized by even lower luminosities (one can note the similarity of  the orbital periods of  
J16479 and J16418,  and that of J16328 to Vela X-1). 
The profiles show more complexity, as more knees are appearing, reflecting different peaks in the DLDs, that is, 
different population of flares. 
Finally, the most extreme SFXTs, 
J08408, J1739,  J17544, and the newly observed J1845.0
 (solid lines) show very complex profiles. 

Considering that both differential/cumulative distributions 
and inactivity duty cycles show that SFXTs are underluminous when compared to HMXBs, 
and that single knee profiles in CLDs  can be understood in terms of wind accretion from an inhomogeneous
medium \citep[see, e.g.][]{Furst2010}, 
we can interpret the differences observed between  
classical systems and SFXTs 
as due to accretion from a structured wind in the former sources 
and to the presence of magnetic/centrifugal gates or a quasi-spherical settling accretion regime in the latter. 

{\bf The king, the power and the ring. }  
Ruler of a small kingdom, the SFXT prototype 
J17544  has been a stimulus, a catalyst, and a continuous 
challenge in the process of understanding SFXTs as a class since 
it was discovered by \inte\ over ten years ago \citep[][]{Sunyaev2003}. 
The optical counterpart in this binary 
\citep[orbital period $P_{\rm orb} = 4.926\pm0.001$\,d, ][]{Clark2009:17544-2619period} 
is quite an ordinary O9Ib star with a mass of 25--28~$M_{\odot}$ \citep{Pellizza2006} 
and located at a distance of 3.6\,kpc \citep{Rahoui2008}. 
The large luminosity swings observed on timescales as short as hours were 
alternatively explained by mechanisms that regulate or inhibit accretion
(\citealt{Stella1986}; \citealt[][]{Grebenev2007}, propeller effect; \citealt[][]{Bozzo2008}, 
magnetic gating). 
In particular, \citet[][]{Bozzo2008} explained them in terms of transitions across the 
magnetic and/or centrifugal barriers. 
In this scenario, the large dynamic range of SFXTs (about five decades) is achieved
with a small variation of the mass loss rate (e.g.\ a factor of
$\sim$5 in fig.\ 3a of \citealt[][]{Bozzo2008}) 
by assuming a spin period $P_{\rm spin}$ in excess of $\sim 1000$~s
and a magnetar-like field ($B \geq 10^{14}$\,G). 
A recent \nustar\ observation \citep[][]{Bhalerao2015:line17544}, however, has revealed a 
cyclotron line at 17\,keV, yielding the first measurement of the magnetic field in a SFXT, at 
$\sim 1.5\times10^{12} $\,G, typical of accreting NSs in HMXBs. 
This set of observations therefore rule out the magnetar nature for the SFXT prototype.

%
\begin{figure}[t]
\begin{center}
\centerline{
\hspace{-0.1cm}
\includegraphics*[angle=0,width=8.9cm]{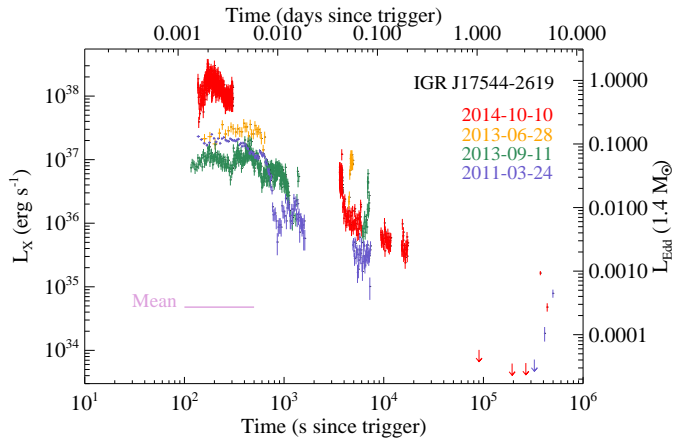}}
\end{center}
\vspace{-1truecm}
\caption{Bolometric X-ray luminosity light curves of 
the brightest outbursts recorded by \sw\ for IGR~J17544$-$2619. 
The giant burst of 2014 October 10 is shown in red, and compared with previous
bright outbursts from this source. 
The horizontal (pink) line marks the average level for this source, obtained from the 
two year monitoring campaign \citep[][see Sect.~\ref{swift10:sec:longterm}]{Romano2011:sfxts_paperVI}, 
while the right-hand y-axis is the standard Eddington luminosity 
for spherical accretion of fully ionized hydrogen 
for a 1.4\,M$_\odot$\ NS.  
Adapted from \citet[][]{Romano2015:17544sb}.  
}
\label{swift10:fig:17544sb}
\end{figure}

On the other hand, the propeller and gating models are still applicable
to fast rotators {($P_{\rm spin} < 10^3$\,s)} even with non-magnetar magnetic
fields like those observed in classical systems.
\citet[][]{Grebenev2010:arXiv1004.0293G} provided a simple equation to estimate
the neutron star spin period at which the magnetic inhibition regime
takes place, 
$$P_{\rm spin}  \approx 4.5 B_{12}^{6/7}\dot{M}_{-5}^{-3/7}v_3^{12/7}P_{10}^{4/7}~{\rm s}, $$
where $B_{12}=B/10^{12}$~G is the magnetic field 
of the neutron star, $\dot{M}_{-5}=\dot{M}_{\rm w}/10^{-5}$~M$_\odot$~yr$^{-1}$
and $v_3=v_{\rm w}/10^3$~km s$^{-1}$ are the mass loss rate and the wind
velocity of the donor star, and $P_{10}=P_{\rm orb}/10$~d
is the orbital period \citep[see also fig.\ 4 of][]{Grebenev2010:arXiv1004.0293G}. 
\citet[][]{Bozzo2008} showed that for sub-magnetar fields
and $\lesssim 100$~s spinning neutron stars,
a $10^5$ luminosity swing would require a much higher increase in the
mass loss rate with respect to the magnetar case (a factor of $\sim 100$
in fig.\ 4 of \citealt[][]{Bozzo2008}).
Such increase would imply that, either the mass loss rate from the 
supergiant stars in SFXTs changes abruptly on short time scales, 
or the local fluctuations in velocity and density of their winds are 
substantially larger than that of classical systems.
In both cases, this would imply a fine tuning for the properties 
of the supergiant star winds in SFXTs compared to other SGXBs.
At present, there is no observational evidence that this should 
be the case, as discussed in \citet[][]{Bozzo2015:underluminous}.

While current investigations concentrate on finding possible mechanisms to inhibit accretion in SFXTs
and to explain their unusually low average X-ray luminosity, J17544 seems to be testing 
our modelling further. 
An exceptionally bright outburst was observed by \sw\ on 2014 October 10 \citep[][]{Romano2015:17544sb},
during which the source reached a peak luminosity of $3\times10^{38}$\,erg\,s$^{-1}$ 
(or a 0.3--10\,keV unabsorbed flux of $1.0\times10^{-7}$ erg cm$^{-2}$ s$^{-1}$, corresponding to 2.1\,Crab).
Tentative evidence for pulsations at a period of $11.6$\,s, and an expanding X-ray halo, or {\it ring}, 
around the source were also found in the XRT data. 
Such a high luminosity (see Fig.~\ref{swift10:fig:17544sb})
not only extends the dynamic range of this source to DR$\sim10^{6}$, a uniquely high value
(by a factor of 10), but also reaches the standard Eddington limit expected for a NS of 
1.4~$M_{\odot}$, thus  challenging, for the first time, the maximum (rather than the minimum) 
theoretical luminosity expected for an SFXT. 
In  \citet[][]{Romano2015:17544sb} we propose that this giant outburst could be caused by 
the formation of a transient accretion disc around the compact object.

\section*{Conclusions}

In the last seven years, \sw\  has contributed many ``firsts'' in the SFXT field,
that I have summarized in this Paper;  most importantly, it has performed the 
first systematic investigation of the soft X-ray long term properties of SFXTs 
with a vey sensitive instrument. 
This has provided us with many clues to help us 
understand SFXT outburst physics; 
it has revised or revolutionized incomplete or over-inferred properties
derived from lower-sensitivity monitorings; 
it has motivated the creation of new models 
(both geometrical and physical) and helped test 
their  applicability. 

\sw\ has consistently surprised us with the unexpected. 
And yet, we are still missing some key ingredients to understand SFXT 
variability, especially when compared with classical systems. 
The current crisis is an excellent motivation to look ``deeper and longer''
and, in this framework, 
\sw\  monitoring programs on SFXTs and classical HMXBs will be crucial.

\section*{Acknowledgements}

I want to thank 
the whole XRT Team, D.N.\ Burrows and J.A.\ Nousek {\it in primis}, for believing we could deliver what we promised;
the BAT Team, S.D.\ Barthelmy and H.A.\ Krimm first in line, for proposing application of the BAT special functions to the SFXT sample,
      and for their invaluable help and support with the BAT and BAT Transient Monitor data; 
the UVOT Team, for never missing a beat (with a cheer).

I also want to thank all current and former collaborators in this endeavour, 
who contributed to the project and who have been teaching me so much:  
E.\ Bozzo, L.\ Ducci, P.\ Esposito, P.A.\ Evans, J.A.\ Kennea, C.\ Guidorzi, V.\ Mangano, S.\ Vercellone;  L.\ Sidoli, 
A.\ Beardmore, M.M.\ Chester,  G.\ Cusumano,  C.\ Ferrigno, C.\ Pagani, K.L.\ Page, D.M.\ Palmer, V.\ La Parola, B.\ Sbarufatti. 
In particular I thank E.\ Bozzo, L.\ Ducci, and P.\ Esposito for a careful reading of the draft. 

I am much in debt to the \sw\ team duty scientists and science planners, 
     truly unsung heroes in my opinion, for providing everything a proposer may possibly need
  (sometimes to the point of anticipating ToOs);  
and of course to Neil Gehrels, PI of this extraordinary and unique discovery machine, 
for running it as a tight but happy ship. I am very, very proud of being part of this crew. 

I thank the referee for comments that helped improve the paper, 
and also acknowledge financial contribution from contract ASI-INAF I/004/11/0.


\end{document}